# Bulk Superconductivity at 14 K in Single Crystals of $Fe_{1+y}Te_xSe_{1-x}$

B. C. Sales, A. S. Sefat, M. A. McGuire, R. Y. Jin, Y. D. Mandrus

*Oak Ridge National Laboratory, Oak Ridge TN 37831, USA*

Y. Mozharivskyj

*Chemistry Department, McMaster University, Hamilton, ON L8S 4M1, Canada*

**Abstract**

Resistivity, magnetic susceptibility and heat capacity measurements are reported for single crystals of $Fe_{1+y}Te_xSe_{1-x}$ grown via a modified Bridgeman method with $0 < y < 0.15$, and x= 1, 0.9, 0.75, 0. 67, 0.55 and 0.5. Although resistivity measurements show traces of superconductivity near 14 K for all x except x=1, only crystals grown with compositions near x=0.5 exhibit bulk superconductivity. The appearance of bulk superconductivity correlates with a reduction in the magnitude of the magnetic susceptibility at room temperature and smaller values of y, the concentration of Fe in the Fe(2) site.

## I. Introduction

The initial discovery[1] of superconductivity at 26 K in fluorine doped samples of LaFeAsO was a surprise since most Fe compounds are magnetic. Within a month of the initial report the transition temperature was raised to as high as 55 K by replacing La with other rare earths such as Ce, Pr, Nd, Sm or Gd. [2-6] Within five months three other crystal structure types of layered Fe compounds were found to be superconducting, including $Ba_{1-x}K_xFe_2As_2$[7], $Li_{1-x}FeAs$[8], and $FeSe$[9]. The common feature of all four tetragonal crystal structures is square planar sheets of Fe in a tetrahedral environment, with a formal valence of $Fe^{+2}$. Although there are structural similarities with the cuprates, superconductivity in the Fe based superconductors appears to arise from the suppression of a metallic magnetic ground state rather than from doping an antiferromagnetic Mott insulator. The itinerant spin-density-wave (SDW) transition in the parent compounds of the Fe based superconductors has several similarities with the SDW transition in Cr metal such as a magnetic susceptibility that increases with increasing temperature above $T_{SDW}$ and a small ordered magnetic moment.[10]

Superconductivity was first reported in polycrystalline FeSe at 8 K[9], followed by pressure studies that indicated a superconducting onset temperature near 27 K for a modest pressure of 15 kbar[11]. Although the initial reports indicated superconductivity only in off-stochiometric $FeSe_{0.82}$ material, careful recent studies indicate that bulk superconductivity (i.e. a substantial heat capacity anomaly near Tc) only occurs for nearly stochiometric FeSe.[12] The superconducting tetragonal phase of FeSe only forms in a narrow temperature (300 °C – 440 °C) and composition window (x =1.01-1.025). [12] This extreme sensitivity to synthesis conditions makes the growth of large single crystals of FeSe difficult. $Fe_{1+y}Te$, however forms in the same tetragonal structure but with $0.06 < y <$ $0.17^{13}$ and can be prepared as large single crystals from the melt[14, 15]. The excess Fe partially occupies the Fe(2) site in the crystal structure (Fig 1a) in between the square planar sheets of Fe. Both experiment[16, 17] and theory[18] indicate that Fe in the Fe(2) site has a local magnetic moment. The interaction between the Fe(2) iron and the more metallic iron in the Fe(1) layer leads to a complicated magnetic structure[16,17] for $Fe_{1.06}Te$ below $T_N$ = 65 K. Theory also suggests that if y = 0, spin-fluctuation mediated superconductivity should be stronger in FeTe than FeSe.[19] Resistivity measurements on polycrystalline alloys[20] of $Fe_{1+y}Te_xSe_{1-x}$ and one single crystal investigation[15] of $Fe_{1.03}Te_{0.7}Se_{0.3}$ give hints of superconductivity with transition temperatures approaching 14 K, but often the resistance does not reach zero, and the magnetic susceptibility screening signal (zfc) is small and does not become negative.

In the present work, we report on the growth and characterization of large single crystals (typical mass about 10 g) of $Fe_{1+y}Te_xSe_{1-x}$, with y as small as allowed by phase stability, and x= 1, 0.9, 0.75, 0.67, 0.55 and 0.5. We find hints of superconductivity for x = 0.9, 0.75, and 0.67, a strong diamagnetic screening signal for x=0.55, and clear evidence of bulk superconductivity for x=0.5.

## II. Experimental Details

Appropriate amounts of Fe pieces (99.99 wt. %), Te shot (99.9999 wt. %) and Se shot (99.9999 wt. %) were loaded into a silica Bridgeman ampoule, evacuated, and sealed. The elements were melted together at 1070 C for 36 h and then cooled in a temperature gradient at rates ranging from 3 to 6 C/h to temperatures ranging from 350-750 C, followed by furnace cooling. Since the silica ampoules often cracked upon cooling, the Bridgeman ampoule was sealed into a second silica ampoule. The starting stochiometry for all of the growths except $Fe_{1.06}Te$ was $FeTe_xSe_{1-x}$ (no excess iron). Typically over half of the resulting boule was a single crystal. All of the crystals could be easily cleaved perpendicular to the *c* axis (Fig 1b, Fig 1c). The phase purity of the crystals were characterized using a Scintag XDS 2000 powder x-ray diffractometer and the chemical composition was measured with a JEOL JSM 840 scanning electron microscope equipped with an EDAX analysis system. Single crystal X-ray diffraction data were collected at room temperature on a STOE IPDSII diffractometer using $MoK_\alpha$ radiation. Transport and heat capacity measurements were performed with a Quantum Design Physical Property Measurement System (PPMS). Electrical leads were attached to the samples using Dupont 4929 silver paste. Magnetic measurements were performed with a Quantum Design (MMPS) SQUID magnetometer.

## III. Results and Discussion

Typical powder x-ray diffraction data from some of the crystals ($Fe_{1+y}Te_xSe_{1-x}$, x= 1, 0.75, 0.67, 0.55, and 0.5) are shown in Fig 2a. All of the patterns are described by the tetragonal PbO structure type with the lattice constants given in Table 1. For the x=0.5 boule, the average composition separates into two tetragonal phases with the tetragonal PbO structure. This separation is evident in the splitting of the (001) line for the x=0.5 composition (Fig 2b). EDAX analysis of different portions of the x=0.5 boule indicate macroscopic stripes (0.1 mm wide stripes) of two phases with *approximate* compositions of $FeTe_{0.53}Se_{0.47}$ (major phase) and $FeTe_{0.35}Se_{0.65}$ (minor phase). A similar phase separation has been reported in polycrystalline samples[20, 21] near this composition range. In spite of the phase separation, neutron diffraction analysis shows that the x=0.5 boule is essentially a single crystal.[22] Subsequent growth experiments have determined that the phase separation observed in the $FeTe_{0.5}Se_{0.5}$ crystal boules can be greatly suppressed or perhaps eliminated using longer soak times at high temperatures. Initial superconducting measurements on these more homogenous samples, however, are virtually identical to the results reported here for the phase separated $FeTe_{0.5}Se_{0.5}$ crystal. EDAX and single crystal refinement analyses of the other crystals $Fe_{1+y}Te_xSe_{1-x}$ (x=1, 0.9, 0.75, 0.67, 0.55) indicate that the Te and Se concentrations are very close to the starting nominal composition. The measured values of y, however, monotonically decrease from y= 0.13 ± 0.02 for x=1, to y = 0 ± 0.02 for x=0.5 (Table 1).

Resistivity data with the current in the *ab* plane are shown in Fig. 3 for crystals cleaved from large boules such as shown in Fig 1. Crystals with 0.5 < x < 0.9 all show indications of superconductivity with an onset drop in the resistivity near 14 K except the crystal with x=0.9 where $T_c$ onset is near 10 K. Note, however, that the resistivity does not reach zero for several of the crystals (x= 0.9, 0.75, 0.67), The smooth drop in the resistivity of $Fe_{1.12}Te_{0.9}Se_{0.1}$ near 35 K is most likely the magnetic/structural transition[16, 17] that occurs at 65 K in $Fe_{1.13}Te$ crystals (see Fig 6). The resistivity of $Fe_{1.13}Te$ is very similar to that reported by several authors.[15, 16]

The magnetic susceptibility at 20 Oe is shown for the x=0.5 crystal (Fig. 4) for both zero field cooled (zfc) and field cooled (fc) measuring conditions, as well as the resistivity data from the same piece. The resistivity reaches zero at about 14 K, consistent with the onset of a large diamagnetic screening signal that reaches full screening ($-1/4\pi$) at 4 K. Only the x=0.55 crystal displayed an equally strong diamagnetic signal expected from a bulk superconductor. The zfc low field susceptibility from the other compositions (x= 0.9, 0.75, 0.67) remained paramagnetic below $T_c$.

Low temperature heat capacity measurements were performed on several $Fe_{1+y}Te_xSe_{1-x}$ crystals with x = 0.67, 0.55 and 0.5 to check for bulk superconductivity. Only the piece from the x=0.5 boule shows a clear heat capacity anomaly at $T_c \approx 14$ K. There is a small heat capacity anomaly near $T_c$ for the x=0.55 sample, but it is not easily visible when the data are plotted on the same scale. The heat capacity data from the x=0.67 crystal shows no evidence of superconductivity and for T< 20 K the data can be accurately described by $\gamma T + BT^3 + CT^5$, with $\gamma$ = 39 mj/mole/K$^2$, and $\Theta_D$ = 174 K. To get an approximate estimate of the electronic contribution to the heat capacity data from the x = 0.5 sample, the data from the x=0.5 and x=0.55 samples were scaled by a few percent to match the data from the x=0.67 sample at 20 K (Fig 5a). The lattice contribution ($BT^3 + CT^5$) from the x=0.67 crystal was then subtracted from the x=0.5 data with the result shown in Fig 5b. Although there is likely a difference in $\gamma$ and the lattice heat capacity between the x=0.67 and x=0.5 compositions, the general shape and magnitude of the data displayed in Fig 5b is typical of a bulk superconducting transition. The activated behavior between 2 and 10 K can be fit to the BCS formula[23] with a gap of about 1.3 $k_BT_c$ somewhat smaller than the expected value of 1.76 $k_BT_c$. We note, however, that not all pieces from the same x=0.5 boule exhibited a clear heat capacity anomaly at $T_c$, indicating that these materials are extremely sensitive to the exact synthesis conditions. This sensitivity was also emphasized in studies of the preparation of polycrystalline FeSe.[12] For FeSe the superconductivity was severely depressed with a small concentration of Fe in the Fe (2) site. This is perhaps not surprising since theory indicates that Fe in the Fe (2) should have a local magnetic moment and hence act as a pair breaker. The value of y, the amount of Fe in the Fe(2) site, decreases for decreasing values of x (Table 1). The high temperature magnetic susceptibility, χ, data from the $Fe_{1+y}Te_xSe_{1-x}$, crystals also suggest that y decreases with decreasing x (more Se) (Fig 6). As the Se content is increased the magnitude of the room temperature susceptibility monotonically decreases from $3.5 \times 10^{-3}$ $cm^3$/mole Fe for x=1 to $1.1 \times 10^{-3}$ $cm^3$/mole Fe for x = 0.5. The room temperature value of χ for other Fe based superconductors such as $LaFeAsO_{0.89}F_{0.11}$ or $Ba(Fe_{1.84}Co_{0.16})As_2$, is typically between $2 - 5 \times 10^{-4}$ $cm^3$/mole. If the occupation of the Fe(2) site can be lowered further through modification of the $Fe_{1+y}Te_xSe_{1-x}$ crystal growth parameters or chemical substitution with a nonmagnetic element, even better superconducting properties could result.

**IV. Summary**

Large single crystals of $Fe_{1+y}Te_xSe_{1-x}$ with x= 1, 0.9, 0.75, 0.67, 0.55, 0.5 and 0 < y < 0.15 were grown using a modified Bridgeman method. Only crystals with x values near 0.5 exhibited bulk superconductivity as evident from resistivity, magnetic susceptibility and heat capacity measurements. The appearance of bulk superconductivity correlates with smaller values of y and the suppression of a magnetic/structural transition between x=0.9 and x=0.75. Relatively large superconducting crystals of these compounds are of interest for neutron scattering and elastic constant experiments aimed at probing the relationship between magnetism, magnetic fluctuations, phonons and superconductivity.

**Acknowledgements**

It is a pleasure to acknowledge enlightening discussions with David Singh, Mark Lumsden, Andrew Christianson, Steve Nagler and Herb Mook as well as the technical assistance of Larry McCollum, Jason Craig, Elder Mellon and Midge Mckinney. This research was supported by the Division of Materials Sciences and Engineering, Office of Basic Energy Sciences, U. S. Department of Energy. Part of this research was performed by Eugene P. Wigner Fellows at ORNL.

Table 1. Chemical composition and lattice constants of $Fe_{1+y}Te_xSe_{1-x}$ single crystals. The lattice constants are determined from single crystal x-ray refinement or from (parentheses) full powder x-ray diffraction patterns using LeBail refinements and the program "FullProf". The differences in composition and lattice constants for crystals prepared from $FeTe_{0.5}Se_{0.5}$ melts reflects the spatial variations of the Te/Se ratios typically found in the resulting boule.

| Starting Melt Composition | Microprobe Composition ($\pm 0.02$) | Single Crystal Refined Comp. | $c$ (Å) | $a$ (Å) |
|---|---|---|---|---|
| $Fe_{1.06}Te$ | $Fe_{1.13}Te$ | $Fe_{1.07}Te$ | 6.284 (6.288) | 3.823 (3.820) |
| $FeTe_{0.9}Se_{0.1}$ | $Fe_{1.12}Te_{0.92}Se_{0.08}$ | $Fe_{1.08}Te_{0.93}Se_{0.07}$ | 6.248 (6.246) | 3.815 (3.813) |
| $FeTe_{0.75}Se_{0.25}$ | $Fe_{1.04}Te_{0.73}Se_{0.27}$ | | (6.163) | (3.809) |
| $FeTe_{0.67}Se_{0.33}$ | $Fe_{1.04}Te_{0.64}Se_{0.36}$ | $Fe_{1.06}Te_{0.68}Se_{0.32}$ | 6.118 (6.109) | 3.803 (3.805) |
| $FeTe_{0.55}Se_{0.45}$ | $Fe_{1.01}Te_{0.54}Se_{046}$ | | (6.050) | (3.798) |
| $FeTe_{0.5}Se_{0.50}$ | $Fe_{0.99}Te_{0.51}Se_{0.49}$ | $Fe_{1.00}Te_{54}Se_{46}$ | 6.069 (6.034) | 3.815 (3.801) |

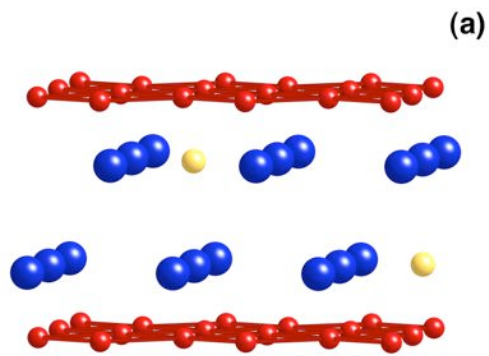

Fig 1a (Color online) Schematic of tetragonal $Fe_{1+y}Te_xSe_{1+x}$ structure. The structure consists of square planar layers of iron (small spheres) and planar layers of Te/Se (large spheres). In most of the crystals a few percent of Fe (medium size spheres) occupies sites in the Se/Te layers.

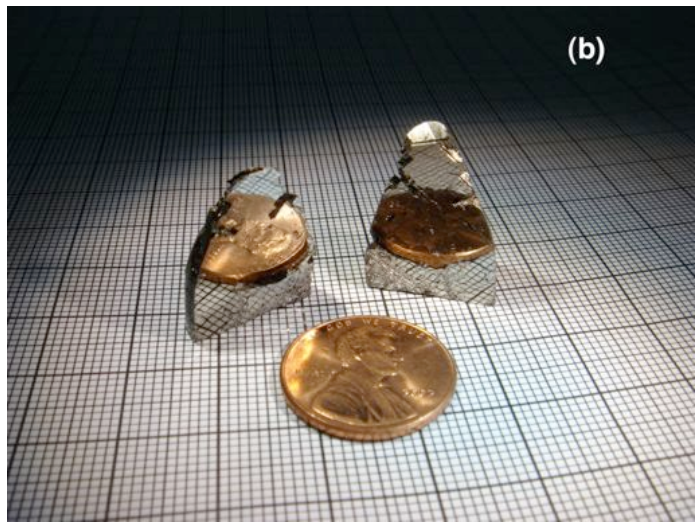

Fig 1b (Color online) Single crystal boule of $Fe_{1.13}Te_{0.73}Se_{0.27}$ cleaved in half with a razor blade. The highly reflective surfaces are perpendicular to the *c* axis. This crystal weighed 17 grams.

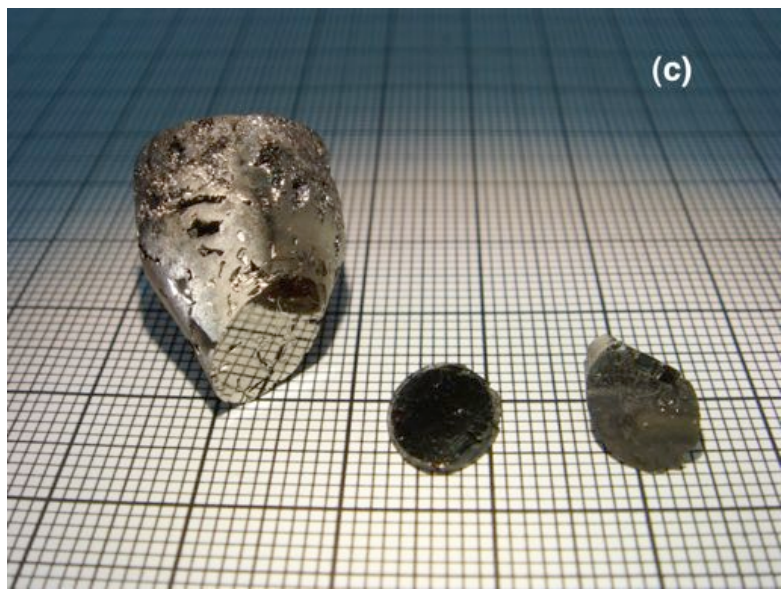

Fig1c. (Color online) Single crystal of $Fe_{1.04}Te_{0.64}Se_{0.36}$ cleaved with a razor blade into three pieces. The highly reflective surfaces are perpendicular to the *c* axis. The crystal weighed about 15 grams.

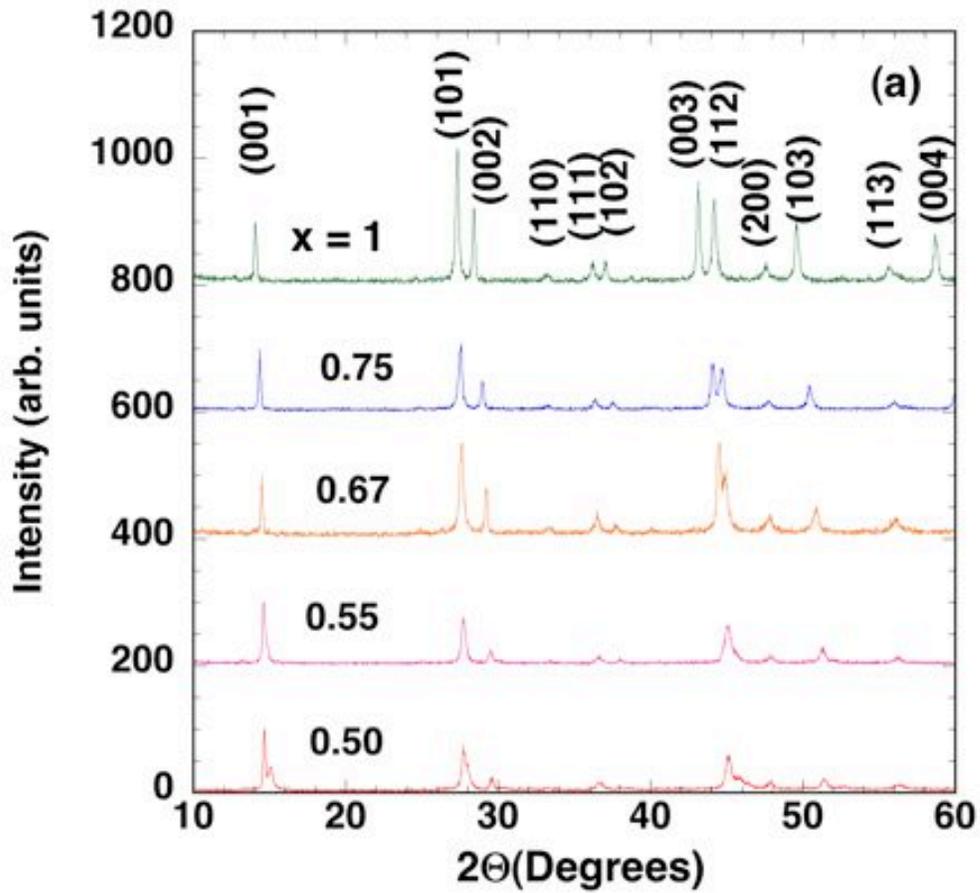

Fig. 2a. (Color online) Powder x-ray diffraction data from $FeTe_xSe_{1-x}$ samples prepared from a small portion of the single crystals such as shown in Fig 1b. The nominal value for x is shown. The actual compositions are given in Table 1.

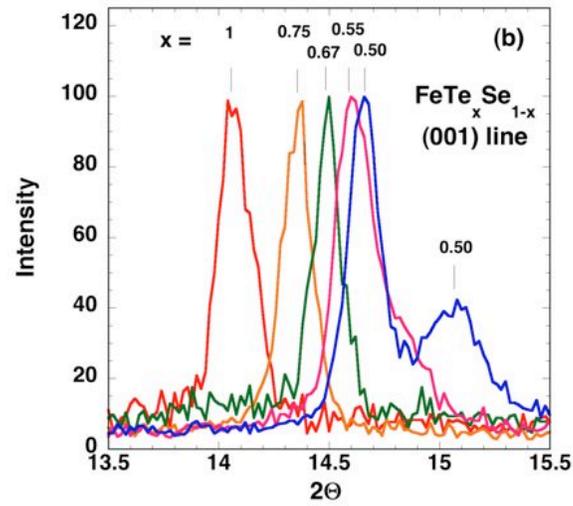

Fig. 2b (Color online) Position of the (001) line with composition. For this x=0.5 crystal there is a macroscopic phase separation into two tetragonal PbO-type phases with different macroscopic compositions.

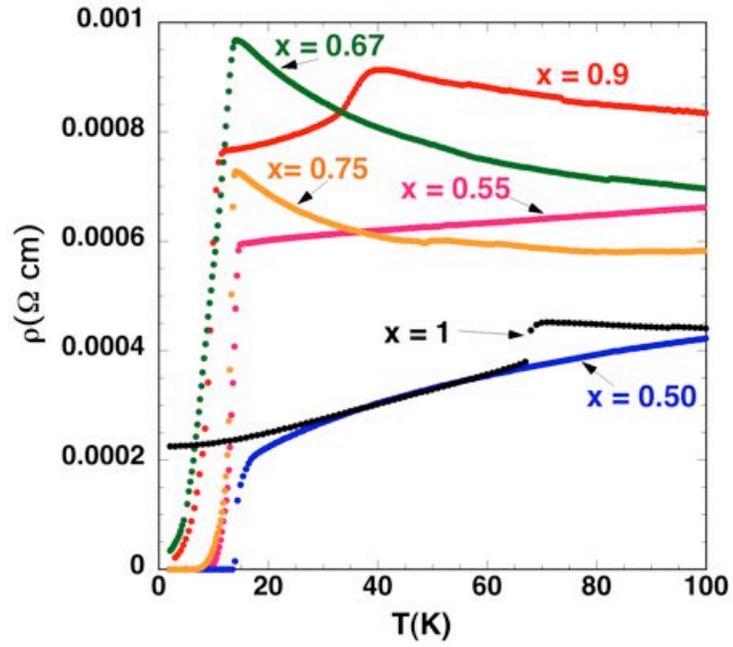

Fig. 3. (Color online) In plane resistivity of $FeTe_xSe_{1-x}$ crystals. Measured compositions are given in Table 1.

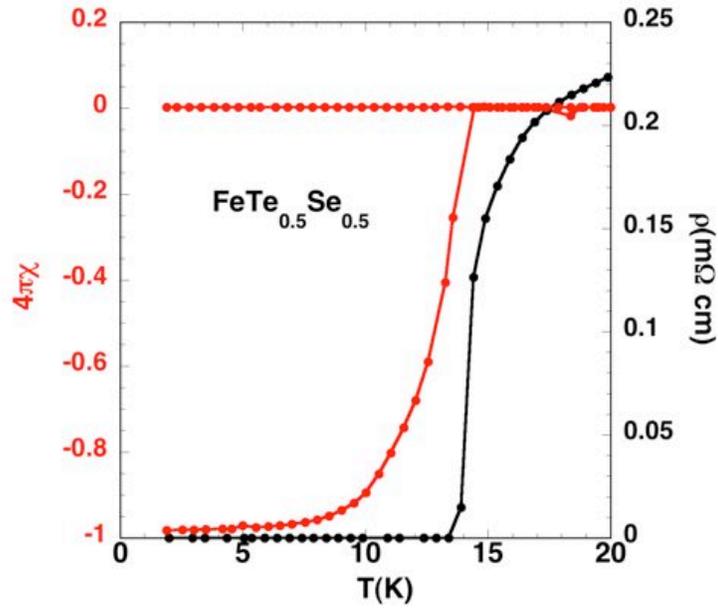

Fig. 4. (Color online) Magnetic susceptibility of a FeTe$_{0.5}$Se$_{0.5}$ crystal measured with H = 20 Oe using zfc and fc protocols. The diamagnetic susceptibility for the zfc data corresponds to complete diamagnetic screening. The resistivity data from the same sample are also shown.

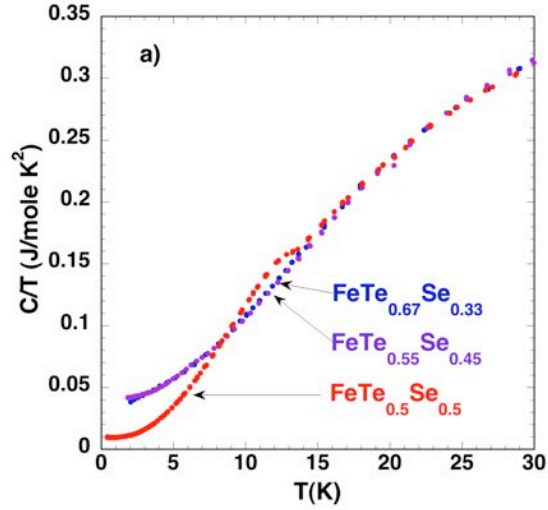

Fig. 5a (Color online) Heat capacity data divided by temperature vs. temperature for three $FeTe_xSe_{1-x}$ crystals. The heat capacity data from the x= 0.5 and x=0.55 crystals have been adjusted by a few percent to match the x=0.67 data at T = 20 K. The heat capacity data for the x=0.67 sample are well described by $\gamma T + BT^3 + CT^5$, with $\gamma$ = 39 mj/mole-$K^2$ and $\Theta_D$ = 174 K. These data were used to subtract the lattice contribution from the x=0.5 data.

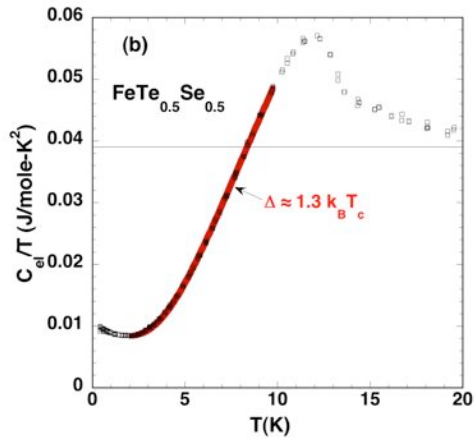

Fig. 5b. (Color online) Estimation of the electronic contribution to the heat capacity of a $FeTe_{0.5}Se_{0.5}$ crystal near and below $T_c \approx$ 14 K. This crystal clearly shows bulk superconductivity. Analysis of the data (line) between 2 and 10 K indicate a value for the gap $\Delta \approx 1.3\ k_BT_c$, some what smaller than the expected BCS value of 1.76.

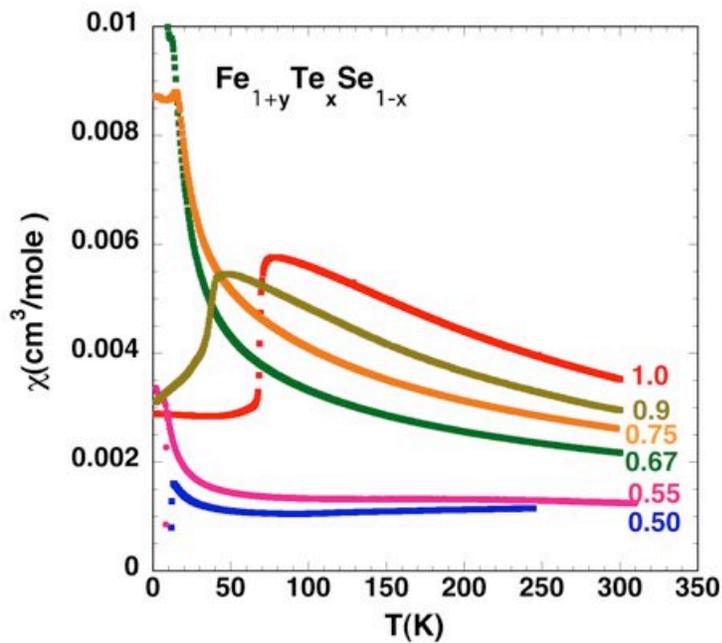

Fig.6. (Color online) Magnetic susceptibility of FeTe$_x$Se$_{1-x}$ crystals with H = 0.1 T and H perpendicular to the *c* axis. Except for the x=0.55 and x=0.5 crystals below the superconducting transition temperatures, the M vs H curves (not shown) are linear from 0 to 7 T. Note that the room temperature susceptibility monotonically decreases with decreasing x.